\documentclass[14pt]{extarticle}

\usepackage{amsfonts, amsmath, euscript, amsthm, amssymb, latexsym, stmaryrd, enumitem}
\usepackage[cp1251]{inputenc}
\usepackage[T2A]{fontenc}
\usepackage[russian]{babel}

\title{Проблема истинности \\ булевых формул с кванторами}
\author{Боков Г.В.}

\newtheorem*{theorem}{Теорема}

\begin{document}

\maketitle

Проблема истинности булевых формул играет ключевую роль в решении комбинаторных проблем. В общем виде она разрешима на детерминированной машине Тьюринга, используя полиномиальное пространство и неограниченное время~\cite{SM73}. В тоже время, для некоторых классов булевых формул эта проблема разрешима за полиномиальное время. Например, проблема истинности булевых формул в конъюнктивной нормальной форме (КНФ), кванторная приставка которых содержит только кванторы существования, разрешима на недетерминированной машине Тьюринга за полиномиальное время~\cite{Coo71,Lev73}. Если же все дизъюнкты в КНФ имеют не более двух литералов~\cite{APT79}, либо имеют не более одного литерала без отрицания~\cite{BKF95}, то проблема их истинности разрешима на детерминированной машине Тьюринга за полиномиальное время. В данной работе будет доказано, что последним свойством обладают все булевы формулы в КНФ.

Пусть $\mathbf{QBF}$ --- множество булевых формул вида $Q_1 x_1 \ldots Q_n x_n \, F$, где $x_1, \ldots, x_n$ --- переменные, $Q_1, \ldots, Q_n \in \{\forall, \exists\}$ --- кванторы, $F = C_1 \wedge \ldots \wedge C_m$, $C_i = x_{i_1}^{a_1} \vee \ldots \vee x_{i_k}^{a_k}$ --- \emph{дизъюнкт} формулы $F$ и $x^1 = x$, $x^0 = \overline{x}$ --- \emph{литералы} переменной $x$. Полагаем $\overline{\overline{x}} = x$. Переменные (их литералы), связанные квантором $Q$, будем называть \emph{$Q$-переменными} (\emph{$Q$-литералами}). Истинность булевых формул определяется стандартным образом: переменные пробегают значения $0$, $1$, а логические связки $\neg$, $\wedge$, $\vee$ интерпретируются как булевы функции отрицания, конъюнкции и дизъюнкции.

Рассмотрим $\Phi = Q_1 x_1 \ldots Q_n x_n \, F \in \mathbf{QBF}$, множество литералов $\mathbf{S}$ и литерал $z$. Пусть $x_i^a \leqslant x_j^b$ всякий раз, когда $Q_i = \exists$ или $i \leq j$; $[\mathbf{S}]_\Phi$ --- множество дизъюнктов $F$, содержащих литералы из $\mathbf{S}$; $\langle\mathbf{S}\rangle_{\Phi,z}$ --- множество $\exists$-литералов $u \neq \overline{z}$ таких, что $z \leqslant u$, $[\{u\}]_\Phi \nsubseteq [\mathbf{S}]_\Phi$ и $[\{\overline{u}\}]_\Phi \subseteq [\mathbf{S}]_\Phi$; $\mathbf{S}_\Phi^0(z) = \{z\}$ и $\mathbf{S}_\Phi^{k+1}(z) = \mathbf{S}_\Phi^k(z) \cup \langle\mathbf{S}_\Phi^k(z)\rangle_{\Phi,z}$ для $k \geq 0$ Положим $\mathbf{S}_\Phi(z) = \mathbf{S}_\Phi^{|\Phi|}(z)$ и $\mathbf{C}_\Phi(z) = [\mathbf{S}_\Phi(z)]_\Phi$, где $|\Phi|$ --- число символов в $\Phi$. Тогда функции $\mathbf{S}_\Phi$, $\mathbf{C}_\Phi$ вычислимы на детерминированной машине Тьюринга за время $O\left(|\Phi|^2\right)$ и для любого литерала $u$ выполнены условиям:
\begin{align*}
  \text{(1)} & \quad u \in \mathbf{S}_\Phi(z) \ \Longrightarrow \ \overline{u} \notin \mathbf{S}_\Phi(z); \\
  \text{(2)} & \quad u \neq z,\ z \leqslant u,\ [\{u\}]_\Phi \subseteq \mathbf{C}_\Phi(z) \ \Longrightarrow \ [\{\overline{u}\}]_\Phi \subseteq \mathbf{C}_\Phi(z).
\end{align*}
Литерал $z$ назовём \emph{избыточным} в $\Phi$, если $[\{\overline{z}\}]_\Phi \subseteq \mathbf{C}_\Phi(z)$. Формулу $\Phi$ назовём \emph{приведённой}, если она не содержит избыточных литералов. Обозначим через $\rho_z(\Phi)$ формулу, полученную из $\Phi$ либо удалением дизъюнктов $\mathbf{C}_\Phi(z)$, если $z$ --- это $\exists$-литерал, либо удалением дизъюнктов $\mathbf{C}_\Phi(\overline{z})$ вместе с удалением литерала $z$ из оставшихся дизъюнктов, если $z$ --- это $\forall$-литерал.

Пусть $z$ избыточный литерал. Тогда $\rho_z(\Phi)$ истинна, если $\Phi$ истинна. Докажем обратное. Пусть $\rho_z(\Phi)$ истинна. Если $\mathbf{S}_\Phi(z) = \left\{x_{i_1}^{a_{i_1}}, \ldots, x_{i_k}^{a_{i_k}}\right\}$, то $i_1, \ldots, i_k$ попарно различны согласно~(1) и не встречаются в $\rho_z(\Phi)$. Если $z$ --- $\exists$-литерал, то все $x_{i_j}$ --- $\exists$-переменные. Тогда, присвоив $x_{i_j}$ значение $a_{i_j}$, все дизъюнкты $\mathbf{C}_\Phi(z)$ станут истинными. Пусть $z$ --- $\forall$-литерал и $i_1 < \ldots < i_k$. Тогда $z = x_{i_1}^{a_{i_1}}$ и все $x_{i_j}$ для $j > 1$ --- $\exists$-переменные. Если $x_{i_1} = a_{i_1}$, то, присвоив $x_{i_j}$ значение $a_{i_j}$, все дизъюнкты $\mathbf{C}_\Phi(\overline{z}) \subseteq \mathbf{C}_\Phi(z)$ станут истинными. Иначе, рассмотрим литералы из $\mathbf{S}_\Phi(\overline{z})$. Тогда, назначив их переменным значения по аналогии с литералами из $\mathbf{S}_\Phi(z)$, все дизъюнкты $\mathbf{C}_\Phi(\overline{z})$ станут истинными. В любом случае, $\Phi$ истинна.

Докажем индукцией по $N_\Phi$ --- числу переменных в $\Phi$, что приведенная формула $\Phi$, содержащая хотя бы один дизъюнкт, ложна. Если $N_\Phi = 0$, то $\Phi$ содержит пустой дизъюнкт и значит ложна. Пусть это верно для всех формул с менее чем $N_\Phi > 0$ переменными, докажем его для $\Phi$. Рассмотрим $Q$-переменную $x$ в $\Phi$ с максимальным номером. Тогда $z \leqslant x^a$ для любого $z$ и $a$. Пусть $[\{x^a\}]_\Phi \setminus [\{\overline{x^a}\}]_\Phi$ состоит из дизъюнктов $C^a_i \vee x^a$, $1 \leq i \leq n_a$, для некоторого $n_a \geq 0$. Тогда $n_0 + n_1 > 0$. Рассмотрим $\Psi$, полученную из $\Phi$ удалением всех дизъюнктов $C^a_i \vee x^a$, добавлением дизъюнктов вида $C^0_i \vee C^1_j$ для всех $i$, $j$ и удалением $Q x$ из кванторной приставки.

Докажем индукцией по $k \geq 0$, что $[\mathbf{S}_\Psi^k(z)]_\Phi \subseteq \mathbf{C}_\Phi(z)$ для любого литерала $z$. Для $k = 0$ верно. Пусть это верно для $k \geq 0$ и $u \in \langle\mathbf{S}_\Psi^k(z)\rangle_{\Psi,z}$. Тогда $u \neq \overline{z}$, $z \leqslant u$, $[\{u\}]_\Psi \nsubseteq [\mathbf{S}_\Psi^k(z)]_\Psi$ и $[\{\overline{u}\}]_\Psi \subseteq [\mathbf{S}_\Psi^k(z)]_\Psi$. Предположим, что $[\{u\}]_\Phi \nsubseteq \mathbf{C}_\Phi(z)$. Согласно~(2), $[\{\overline{u}\}]_\Phi \nsubseteq \mathbf{C}_\Phi(z)$. Пусть $C \in [\{\overline{u}\}]_\Phi \setminus \mathbf{C}_\Phi(z)$. Если $C \in [\{\overline{u}\}]_\Psi$, то $C \in [\mathbf{S}_\Psi^k(z)]_\Psi$ и тогда $C \in [\mathbf{S}_\Psi^k(z)]_\Phi \subseteq \mathbf{C}_\Phi(z)$, что невозможно. Иначе, $C = C^a_{i_a} \vee x^a$ для некоторых $a$, $i_a$ и, следовательно, $C^0_{i_0} \vee C^1_{i_1} \in [\{\overline{u}\}]_\Psi \subseteq [\mathbf{S}_\Psi^k(z)]_\Psi$ для всех $i_{\overline{a}}$. Поэтому либо $C \in [\mathbf{S}_\Psi^k(z)]_\Phi \subseteq \mathbf{C}_\Phi(z)$, либо $[\{x^{\overline{a}}\}]_\Phi \subseteq [\mathbf{S}_\Psi^k(z)]_\Phi \subseteq \mathbf{C}_\Phi(z)$ и тогда $C \in [\{x^a\}]_\Phi \subseteq \mathbf{C}_\Phi(z)$ согласно~(2), что невозможно. Значит, $[\{u\}]_\Phi \subseteq \mathbf{C}_\Phi(z)$ и $[\mathbf{S}_\Psi^{k+1}(z)]_\Phi \subseteq \mathbf{C}_\Phi(z)$. Таким образом, $[\mathbf{S}_\Psi(z)]_\Phi \subseteq \mathbf{C}_\Phi(z)$ для любого литерала $z$.

Предположим, что $[\{\overline{z}\}]_\Psi \subseteq \mathbf{C}_\Psi(z)$. Тогда $[\{\overline{z}\}]_\Phi \nsubseteq \mathbf{C}_\Phi(z)$. Пусть $C \in [\{\overline{z}\}]_\Phi \setminus \mathbf{C}_\Phi(z)$. Если $C \in [\{\overline{z}\}]_\Psi$, то $C \in [\mathbf{S}_\Psi(z)]_\Psi$ и тогда $C \in [\mathbf{S}_\Psi(z)]_\Phi \subseteq \mathbf{C}_\Phi(z)$, что невозможно. Иначе, $C = C^a_{i_a} \vee x^a$ для некоторых $a$, $i_a$ и, следовательно, $C^0_{i_0} \vee C^1_{i_1} \in [\{\overline{z}\}]_\Psi \subseteq [\mathbf{S}_\Psi(z)]_\Psi$ для всех $i_{\overline{a}}$. Поэтому либо $C \in [\mathbf{S}_\Psi(z)]_\Phi \subseteq \mathbf{C}_\Phi(z)$, либо $[\{x^{\overline{a}}\}]_\Phi \subseteq [\mathbf{S}_\Psi(z)]_\Phi \subseteq \mathbf{C}_\Phi(z)$ и тогда $C \in [\{x^a\}]_\Phi \subseteq \mathbf{C}_\Phi(z)$ согласно~(2), что невозможно. Значит, $\Psi$ приведенная формула и по предположению индукции $\Psi$ ложна. Так как истинность $\Phi$ влечет истинность $\Psi$, формула $\Phi$ также ложна.

Теперь заметим, что $|\rho_z(\Phi)| < |\Phi|$ для любого избыточного литерала $z$. Поэтому не более чем за $O(|\Phi|)$ шагов формулу $\Phi$ можно свести к такой приведенной формуле $\Psi$, что $\Phi$ является истинной тогда и только тогда, когда $\Psi$ не содержит дизъюнктов. Поскольку каждый шаг может быть выполнен на детерминированной машине Тьюринга за время $O\left(|\Phi|^3\right)$, справедлива теорема.

\begin{theorem} \label{T:1}
Существует детерминированный алгоритм, который по $\Phi \in \mathbf{QBF}$ остановится за $O\left(|\Phi|^4\right)$ шагов и выдаст 1 если $\Phi$ истинна, 0 иначе.
\end{theorem}

Пусть $\mathbf{P}$ ($\mathbf{PSPACE}$) --- класс всех задач, решаемых на детерминированной машине Тьюринга за полиномиальное время (используя полиномиальное пространство и неограниченное время). Тогда $\mathbf{P} \subseteq \mathbf{PSPACE}$. Поскольку любую задачу из $\mathbf{PSPACE}$ можно эффективно свести к проблеме истинности формул из $\mathbf{QBF}$~\cite{SM73}, верно равенство $\mathbf{P} = \mathbf{PSPACE}$.

\newpage

УДК 510.52

Боков Григорий Владимирович

МГУ имени М.~В.~Ломоносова

Механико-математический факультет

Кафедра математической теории интеллектуальных систем

Лаборатория математических проблем искусственного интеллекта

Адрес: 119991, Москва, ГСП-1, Ленинские горы, д. 1, механико-математический факультет, кафедра математической теории интеллектуальных систем, лаборатория математических проблем искусственного интеллекта.

Служ. тел.: +74959395421

Моб. тел.: +79162263128

Email: bokov@intsys.msu.ru

\end{document}